# Horizontal and vertical energy fluxes of ocean surface waves and their derivation from spaceborne altimeter measurements

Paul A. Hwang

*Remote Sensing Division, Naval Research Laboratory, Washington DC*

**Running title**: Space remote sensing of energy fluxes

Recent research shows that the surface wave energy dissipation, which is the vertical energy flux across the air-sea interface, can be calculated as the product of air density, reference wind speed cubed and an energy transfer coefficient determined by the dimensionless parameters made of wind speed, significant wave height and dominant wave period. In a similar way, the horizontal wave energy flux of wind generated waves can be represented by the same dimensionless wind and wave parameters. Satellite altimeters routinely report reference wind speed and significant wave height. An algorithm to derive the characteristic wave period of ocean waves in the altimeter footprint using the similarity properties of ocean wind and waves is described. The vertical and horizontal energy fluxes derived from the satellite altimeter are in very good agreement with the estimation from ocean buoy measurements in four geography locations with significantly different wind and wave climates. The vertical energy flux follows closely the cubic wind speed dependence, reflecting the dominance of short wave contribution in wave generation and breaking dissipation. The wind speed dependence of horizontal energy flux is much weaker especially in mild to moderate wind speed, reflecting its dominance by long swell component. Application of the energy flux parameterization functions to satellite altimeter measurements offers an efficient method of estimating the air-sea exchange and ocean energy budget in global scale. Such data are extremely difficult to acquire using other means.

**1. Introduction**

Satellite remote sensing offers an efficient way of monitoring global and regional earth environments. Spaceborne altimeters, following their many generations of development, have provided high-quality wind speed and wave height data over the world's oceans with an unusually high spatial density along the satellite ground tracks. For example, TOPEX/POSEIDON (hereafter referred to as TP) reports wind and wave information at one-second intervals and produces measurements at approximately every 7 km along the satellite groundtrack. The spacing between neighboring tracks is nominally 316 km at the equator and much smaller at higher latitudes (127 revolutions per repeat cycle). Verifications with collocated and simultaneous ocean buoy data have shown that the altimeter derived significant wave height agrees with buoy data to within 0.15 m and the wind speed accuracy is approximately 1.7 m/s globally and 1.2 m/s regionally [e.g., *Cotton and Carter*, 1994; *Ebuchi and Kawamura*, 1994; *Freilich and Challenor*, 1994; *Gower*, 1996; *Hwang et al.*, 1998]. Furthermore, the wave height and wind speed can be used to derive a characteristic wave period using empirical functions correlating the three quantities: wave height, wave period and wind speed. The characteristic period calculated from altimeter output correlates very well with the average or peak wave period measured by in situ buoys in the Gulf of Mexico [*Hwang et al.*, 1998]. *Davies et al.* [1998] take a semi-theoretical approach relating the backscattering cross section and the directional properties of wave spectrum to obtain a characteristic wave period from altimeter data. The wave age dependence on the derived wave period is discussed and the comparison with buoy measurement is very good. More recently, *Gommenginger et al.* [2003] make use of the property that the altimeter backscattering cross section is inversely proportional to the mean square slope of the ocean surface. Assuming that the mean square slope can be represented by the ratio of the significant wave height and dominant wavelength, they establish empirical correlation functions from collocated and simultaneous altimeter and ocean buoy measurements of the peak, mean and zero crossing wave periods ($T_p$, $T_m$ and $T_z$); the latter two quantities are derived, respectively, from the first and second moments of the wave spectrum. Their results are in very good agreement with ocean buoy data. On the NDBC web site, $T_p$ and $T_z$ are routinely reported [*Earle*, 1996; IAHR, 1989], the latter quantity is called average wave period on the NDBC web site and denoted by the variable $T_a$ in this paper.

The feasibility of deriving wave period from altimeter data is quite interesting because the energy transfer across the air-sea interface can be evaluated from the surface wave energy dissipation, $\varepsilon_b$, which can be calculated from simultaneous measurements of reference wind speed (neutral wind speed at 10 m elevation), $U_{10}$, significant wave height, $H_s$, and dominant wave period at the spectral peak, $T_p$ [*Hwang and Sletten*, 2008]. Very good agreement is found in the comparison of the energy dissipation calculation using the parameterization function with field measurements of *Felizardo and Melville* [1995], *Terray et al.* [1996], *Hanson and Phillips* [1999] and *Gemmrich and Farmer* [2004]. Another quantity that is of great interest is the horizontal wave energy flux across a vertical plane, which can be used to evaluate the "energy reserve" of an ocean wave field. This is usually represented in its vertically integrated form, and can be expressed as the product of wave energy and group velocity, the latter is determined uniquely by the wave period in deep water wave condition.

In this paper, the application of the vertical and horizontal energy flux parameterization functions based on the spaceborne altimeter measurements is illustrated. Two issues are addressed prior to this application: (a) The characteristic wave period presented in *Hwang et al*. [1998] can be further refined to correct for some apparent wave age dependent discrepancies with in situ measurement. Collocated and simultaneous altimeter and ocean buoy data sets from Gulf of Alaska and Bering Sea (GOA), Gulf of Mexico (GOM), Hawaiian Islands (HAW), and the equatorial region (EQU) are compiled to establish the refinement algorithms for a broad range of environmental conditions (Section 2). (b) The parameterization functions of the energy fluxes are derived on the assumption of steady wind forcing and fetch- or duration-limited wave growth. The applicability issue of the wind sea formulation to mixed sea conditions needs to be addressed. This is done in both Sections 2 and 3 with the wind and wave measurements derived from the ocean buoys described above. Comparisons of the vertical and horizontal energy fluxes derived from spaceborne altimeter and ocean buoys are then described. A summary is presented in Section 4.

## 2. Derivation of significant wave period from altimeter data

### 2.1. Empirical similarity relation

*Hwang et al.* [1998] describe a procedure to calculate a characteristic wave period, $T_0$, from altimeter output of significant wave height, $H_s$, and wind speed, $U_{10}$, using empirical correlations between wave period, wave height and wind speed established from earlier wave research [e.g., *Hasselmann et al.,* 1973; *Toba*, 1978]. Fig. 1 shows an example of the close correlation among the three variables presented in different forms of dimensionless wave height and wave frequency. In Fig. 1a, $\eta_*\left(\omega_*\right)$ is presented, where $\eta_* = \eta_{rms}^2 g^2 / U_{10}^4$ with $\eta_{rms}^2$ the variance of surface displacement and $g$ the gravitational acceleration; and $\omega_* = \omega_r U_{10} / g$ with either peak wave frequency $\omega_p$ or average wave frequency $\omega_a$ serving as the reference frequency $\omega_r$. The data clouds are the GOA measurements (567 data points) shown in light colors in the background with the size of the plotting symbols proportional to the data density. The dark colored symbols are the bin average with error bars representing one standard deviation. The smooth curves represent the similarity properties derived from analyses of wind-generated waves under fetch-limited growth conditions expressed as a power law function [*Hwang*, 2006]

$$\eta_* = R\omega_*^{\,r}. \tag{1}$$

The solid curve is for the second order fitting analysis with the coefficient and exponent $R$ and $r$ varying with the stage of wave growth, that is, they vary with dimensionless duration, fetch or reference frequency. Lookup tables for the coefficients and exponents of the second order fitting analysis of the wind generated wave growth functions have been given in *Hwang and Sletten* [2008, Table 1]. As shown in the table, the slope of $r$ with respect to $\omega_*$ approaches zero at $\omega_* \approx 0.7$ and the computation of the growth curve is valid only for $\omega_* \geq \sim 0.7$. An approximation of $R$ and $r$ can be written as

$$R = \exp(-6.1384)\omega_*^{0.6102\ln\omega_*}, \quad r = -2.4019 - 1.2204\ln\omega_*. \tag{2}$$

The dashed curve is for the first order fitting analysis, which yields constant proportionality coefficient and exponent, $R_1 = 2.94\times10^{-3}$ and $r_1 = -3.42$.

For wind generated wave data, $\omega_p$ is almost always used as the reference wave frequency due to its correspondence to the wave frequency component of maximum height and energy, thus it is of primal importance in ocean engineering applications. Experimental observations indicate that wind generation of surface waves becomes ineffective when wave phase speed exceeds about $1.25U_{10}$, corresponding to $\omega_* < 0.8$ [*Pierson and Moskowitz*, 1964]. The data shown in Fig. 1a indicate that even in the mixed sea conditions in the open ocean, wave properties in the region with $\omega_* \geq \sim 0.8$ can be described by the windsea growth function (1) with $\omega_p$ as the reference wave frequency. Statistically, $\omega_d$ is about $1.3\omega_p$ [*Hwang et al.*, 1998, Fig. 12d], and the data of $\eta_*\left(\omega_*\right)$ shift to the right (or equivalently, upward) as shown in Fig. 1a. The ratio of upshift is approximately 1.6 times based on the data sets assembled from the four geophysical regions in this paper (see also Fig. 7 later). Figs. 1b and 1c show the results with slightly different expressions of the dimensionless wave height and wave period together with the empirical curves given by *Hasselmann et al.* [1973] and *Toba* [1978], as described in *Hwang et al.* [1998]. The first order fitted equation (1) can be rewritten with the variation of dimensionless parameters:

$$\frac{U_{10}}{gT_0} = 6.517 \times 10^{-2} \left( \frac{U_{10}}{gH_s} \right)^{0.5843} ,$$

(3)

where $T_0$ is a characteristic wave period. Using a slightly variation of (3) *Hwang et al.* [1998] notice that $T_0$ obtained from TP wind speed and wave height is neither $T_p$ nor $T_a$ but a quantity in between. Subsequent analysis reveals that some discrepancies with buoy data are wave age dependent, pointing a way for further refinement. The refinement algorithm will be presented in Subsection 2.3 after the description of data sets used in the analysis.

## 2.2. Environmental characteristics of datasets from four regions

Collocated and simultaneous wind and wave data from NDBC (National Data Buoy Center) buoys and TP groundtracks are collected from four regions with distinctive wind and wave conditions. The maximum time and space differences between buoy locations and altimeter footprints are set to be 0.5 h and 100 km. The detailed information on merging buoy and altimeter data sets has been presented in *Hwang et al.* [1998] and will not be repeated here. Table 1 lists the buoy stations and satellite tracks used. The locations of the buoys are marked on the map in Fig. 2. Most of the buoys are operational over the 7 years of the TP data (1992-1999) except in the equatorial region, where only one year's measurements are available at the time of data compilation.

The probability distribution functions (pdf) of wind speed, significant wave height, average and peak wave periods, and dimensionless wave frequencies referenced to average and peak wave periods of the four regions are shown in Fig. 3. The corresponding statistics of mean and standard deviation are listed in Table 2 for reference. The coarse resolution of peak wave period, $T_p$, reported in the NDBC buoy data is clearly seen in its pdf (Fig. 3c) therefore $T_a$ is chosen as the reference wave period in the development of altimeter algorithm. As expected, the GOA region is characterized by high sea states with strong winds and high waves. Due to its enclosed nature, GOM is typical of a low sea state region with low wind speed, wave height and short wave period. In terms of the wave age parameter (inversely proportional to $\omega_*$), the two regions are quite similar. The equatorial region is influenced by background swell, reflected by the large wave period and relatively low wind speed. The mean wave age of this area is the largest of the four regions. Swell condition in the Hawaiian region is also severe. The data collection of these four regions, therefore, represents a broad coverage of the ocean wind and wave conditions.

### 2.3. Derivation of wave period from spaceborne altimeter

Following the approach of *Hwang et al.* [1988], (3) is applied to the collocated data described above. The first level solution of the wave period

$$T_0 = \frac{U_{10}}{6.517 \times 10^{-2} \, g} \left( \frac{U_{10}^{\,2}}{g H_s} \right)^{-0.5842}, \qquad (4)$$

derived from TP and the measured $T_a$ from buoys are highly correlated. Fig. 4a shows the result with the GOA data as an illustration. Results from application to the other regions are similar (see also *Hwang et al.*, [1998, (Fig. 14c)]). While highly correlated, $T_0$ from TP is clearly larger than $T_a$ from buoy. A linear regression is performed to remove the trend of over estimation,

$$T_1 = a_1 T_0 + a_0,$$  (5)

where $T_1$ is the first iteration wave period, $a_1$=6.746×10$^{-1}$ and $a_0$=1.679 are empirically determined with polynomial fitting to the GOA data using the buoy $T_a$ in place of $T_1$ in (5). The agreement between $T_1$ and the buoy $T_a$ is improved considerably (Fig. 5a). There remains a mild dependence on wave age as illustrated by the two data subgroups with different ranges of wave age and plotted using different symbols in Fig. 4a. The trend can also be detected when the derived wave period is plotted against the wave age (Fig. 4c). Because the purpose of the algorithm is to derived wave period, and then the energy flux, from altimeter output, a surrogate wave age, $A_1$, is calculated with the altimeter-derived $T_1$,

$$A_1 = \frac{g T_1}{2\pi U_{10}}.$$  (6)

Fig. 4b shows the comparison of $A_1$ with the buoy wave age, the two are proportional to each other in general. The wave age dependence is then removed using a second order polynomial function

$$\frac{T_a}{T_1} = b_2 A_1^2 + b_1 A_1 + b_0.$$  (7)

The coefficients $b_2$=-3.377×10$^{-2}$, $b_1$=2.254×10$^{-1}$, and $b_0$=7.564×10$^{-1}$ are again empirically determined from the GOA data set.

After the procedure of linear regression (5) and wave age correction (7), the wave period derived from TP is quite comparable to the buoy $T_a$ (Fig. 5b) and the resulting wave age bias of $T_a$ from TP is mostly removed (Fig. 5c). The algorithm is applied to the other three geophysical regions to obtain the average wave period from altimeter wind speed and wave height. The statistics of bias ($B$), orthogonal regression coefficient ($c$, see *Hwang et al.* [1998, Appendix]), rms difference ($D$), and correlation coefficient ($Q$) between the TP and buoy $T_a$ are listed in Table 3. From here on, $T_a$ is used for the average wave period derived from altimeter and wave buoy. If distinction between the two is needed, subscript 'T" or 'TOPEX' for TP and 'B' or 'Buoy' for buoy will be added.

Fig. 6 compares the average wave period derived from the algorithms described here and in *Gommenginger et al.* [2003]. As mentioned in the Introduction, NDBC reports $T_a$ based on the second moment of wave spectrum [*Earle*, 1996], which is $T_z$ in *Gommenginger et al.* [2003]. Both algorithms perform well, the present method yields somewhat better agreement with the buoy data based on the application to the four geophysical regions in the compiled data. Table 4 lists the relevant statistics of bias, orthogonal linear regression coefficient, root mean square difference, correlation coefficient, and normalized root mean square difference ($D_N$) between the buoy $T_a$ and the average wave period derived from each algorithm.

## 3. Energy flux parameterization

### 3.1. Horizontal energy flux

The magnitude of the horizontal energy flux of surface waves across a vertical plane, usually represented by its vertically integrated form, can be computed from the product of the surface wave energy density and the group velocity, $\varepsilon_h = Ec_g$, where $E = \rho_w g \eta_{rms}^2$ is the total wave energy (sum of potential and kinetic energy), $\rho_w$ water density ($\rho_w$=1030 kg/m$^3$ used in all calculations in this paper) and $c_g$ wave group speed [e.g., *Phillips*, 1977; *Dean and Dalrymple*, 1991]. In dimensionless form, this can be written as

$$\frac{\varepsilon_h}{\rho_w U_{10}^5} = \alpha_h = \frac{1}{2}\omega_*^{-1}\eta_*, \qquad (8)$$

where $\alpha_h$ is the nondimensional coefficient of horizontal energy transfer, uniquely determined by the dimensionless properties of characteristic wave frequency and wave height as shown in (8). As obvious from inspecting Fig. 1a, the numerical value of $\alpha_h$ may differ depending on whether $T_p$ or $T_a$ is used as the characteristic wave period. Figs. 7a and 7b show $\alpha_h$ computed with the two wave periods based on buoy data in the four geophysical regions. Using the windsea wave generation curves as reference, a factor of 0.6 is applied to the calculation with $T_a$ as the reference wave period to account for the observation that the magnitude of $\eta_*(\omega_{a*})$ is larger than $\eta_*(\omega_{p*})$ by a factor of about 1.6 for the same numerical value of $\omega_{a*}$ and $\omega_{p*}$ (Figs. 1a, 7c and 7d). Fig. 8a shows $\alpha_h$ computed with the TP measurements applying the correction factor of 0.6, because the derived wave period is the equivalent $T_a$. Fig. 8b displays the computed $\varepsilon_h$ using the buoy and TP wind and wave data. The two sets of horizontal energy flux estimates are in good agreement.

For ocean applications, sometimes wave information is not available. It is of interest to know whether the energy flux is expressible as a function of wind speed alone. Fig. 8c shows $\alpha_h$ vs. $U_{10}$. The result is suggestive of a trend of $\alpha_h$ approaching asymptotically to a constant value of about $4\times10^{-4}$ toward high wind speed. This value corresponds to $\alpha_h$ at $\omega_{a*}$ about 1.3 (Fig. 8a) and suggests that in the open ocean, the sea state of surface waves tends to gravitate to $\omega_{a*}$ near 1.3 at high wind speeds. Fig. 8d shows $\varepsilon_h$ as a function of $U_{10}$. At low and moderate wind speed, the horizontal energy flux is almost constant and independent of wind speed, consistent with our expectation that the swell influence becomes dominant toward lower wind condition because the wave energy field is mainly contributed by higher and longer waves, that is, $\varepsilon_h = Ec_g$. The background level of horizontal energy flux in Pacific Ocean is around 50 to 100 kW/m. In the Gulf of Mexico, it is about 6 kW/m, roughly one order of magnitude smaller. At high wind speed, $\varepsilon_h \approx 4 \times 10^{-4} \rho_w U_{10}^5$ or $\varepsilon_h \approx 0.4 U_{10}^5$ represents a reasonable asymptote of the horizontal energy flux dependence on wind speed, where $\varepsilon_h$ is in W/m and $U_{10}$ in m/s; but the applicable wind speed with the asymptotic equation exceeds about 10 m/s in closed water bodies such as the Gulf of Mexico and well over 15 m/s in the open ocean.

Interestingly, while GOA has the highest range in wind speed and wave height, for modest and low wind conditions, it has the lowest $\varepsilon_h$ among the three regions in the Pacific Ocean studies here, falling below even EQU with low wind and mild swell (Fig. 3)! This is good news for wave energy extraction – the ocean wave energy is dominated by the swell portion of the wave spectrum and substantial reserve is available even in calm and temperate regions in the open ocean. This is significantly different from the wind energy. In future expansion of human activities from land to ocean, energy supply can be obtained locally in most regions of the ocean.

### 3.2. Vertical energy flux

The vertical energy flux at the air-sea interface represents the energy input from atmosphere to the ocean with surface waves serving as the transfer medium. From decades of wind wave growth research, the atmospheric input to the wave field is approximately equal to the wave energy dissipation [e.g., *Hasselmann et al.*, 1973; *Phillips*, 1985; *Donelan*, 1998; *Hwang and Sletten*, 2008]. The parameterization function of surface wave energy dissipation is given by *Hwang and Sletten* [2008]

$$\varepsilon_v = \alpha_v \rho_a U_{10}^3, \text{ with } \alpha_v = 0.20\omega_*^{3.3}\eta_* , \qquad (9)$$

where $\alpha_v$ is the vertical energy transfer coefficient relating the energy dissipation with wind speed and $\rho_a$ the density of air ($\rho_a$=1.2 kg/m³ used in all calculations in this paper). *Hwang and Sletten* [2008] show that for wind events in the ocean with wind duration longer than one hour ($\omega_*$ is generally less than 4), the numerical value of $\alpha_v$ for practical applications is $(3.7\sim5.7)\times10^{-4}$. For more general conditions, say $\omega_*<10$, the range of $\alpha_v$ is about $(2\sim6)\times10^{-4}$ [*Hwang and Sletten*, 2008, Fig. 3a]. Computations using the parameterization function (9) are in very good agreement with field measurements [*Felizardo and Melville*, 1995; *Terray et al.*, 1996; *Hanson and Phillips*, 1999; *Gemmrich and Farmer*, 2004).

Fig. 9 shows $\alpha_v$ calculated with the dimensionless parameterization equation using the wind and wave output from buoy (Fig. 9a) and TP (Fig. 9b). A factor of 0.6 is applied to the calculation to account for the observation that the magnitude of $\eta_*(\omega_{u*})$ is larger than $\eta_*(\omega_{p*})$ by a factor of about 1.6, as mentioned earlier. Fig. 9c illustrates the close resemblance of the vertical energy fluxes computed from buoy and TP data. The vertical energy flux is strongly dependent on wind speed, reflecting its dominance by short scale waves. The dominance by short waves in air-sea energy transfer is generally accepted for the case of wind generation [e.g., *Phillips*, 1985; *Donelan*, 1988; *Hwang and Sletten*, 2008]. For wave breaking, recent field data provide ample evidence showing that the breaking velocity and length scales are concentrated in short waves with wave period on the order of about 1 to 2 s [*Hwang and Wang*, 2004; *Hwang et al.*, 2008a, b; *Gemmrich et al.*, 2008].

## 4. Summary and conclusions

In this paper, a procedure to derive the energy flux from altimeter output of significant wave height and wind speed. The algorithm is based on the robust correlation between two dimensionless parameters, $U_{10}/gT_a$ and $U_{10}^2/gH_s$, or equivalently, the dimensionless frequency $\omega_*$ and variance $\eta_*$, established from earlier ocean wave research [e.g., *Hasslemann et al.*, 1973; *Toba*, 1978; *Donelan et al.,* 1985; *Hwang*, 2006]. The resulting characteristic wave period obtained with this approach shows close correlation with the peak or average wave period measured by in situ buoys [*Hwang et al.*, 1998]. The wave period derived from the altimeter source using the algorithm described in *Hwang et al.* [1998] has a magnitude in between the peak and average wave periods routinely reported by NDBC. Using buoy $T_a$ as reference, a linear regression (3) is applied to remove the overestimation. After the linear regression, there remains a mild wave age dependence in the discrepancy between TP and buoy wave periods (Fig. 4c), which can be removed by a second order polynomial function (5). The wave period computed from TP wind speed and wave height following these two additional steps is in good agreement with the average wave period measured by in situ buoys (Fig. 5b and Table 3). With this wave period algorithm, the vertical and horizontal energy fluxes of a wave field can be computed from TP output of wind speed and wave height. The horizontal and vertical energy fluxes derived from spaceborne altimeter are in excellent agreement with those from in situ buoys (Figs. 8b and 9c). For the vertical energy flux, which is dictated by shorter scale waves, it follows closely the cubic wind speed dependence. A practical approximation is $\varepsilon_v \approx (2.4 \sim 7.2) \times 10^{-4} U_{10}^3$, with $\varepsilon_v$ in W/m$^2$ and $U_{10}$ in m/s. For the horizontal energy flux, which is dominated by long swell, the dependence on wind speed is much weaker. Substantial ambient level of horizontal energy flux exists in mild wind and temperate regions of the ocean. In the Pacific Ocean, the ambient level is about 50~100 kW/m, and in the Gulf of Mexico, it is about 6 kW/m. The result highlights the significant role of ocean surface waves in global energy transfer.

For future expansion of human activities from land to ocean, plenty of energy from surface waves can be derived locally in most part of the ocean.

**Acknowledgments.** This work is sponsored by the Office of Naval Research (Naval Research Laboratory PE62435N and PE61153N). The TP data are provided by Gregg Jacobs and Bill Teague. Buoy wind and wave data are provided by NDBC. David Wang contributed in merging the TP and buoy data. (NRL contribution JA/7260-08-xxxx).

### References

Cotton, P. D., and Carter, D. J. T. (1994), Cross calibration of TOPEX, ERS-1, and Geosat wave heights, *J. Geophys. Res.*, *99*, 25025-25033.

Davies, C. G., Challenor, P. G., and Cotton, P. D. (1998). Measurements of wave period from radar altimeter, in B. L. Edge and J. M. Hemsley (Eds.) *Ocean Wave Measurement and Analysis*, 809-818, Reston: ASCE.

Dean, R. G., and R. A. Dalrymple (1991), *Water wave mechanics for engineers and scientists*, World Scientific Publ., 353pp.

Donelan, M. A. (1998), Air-water exchange processes, in *Physical Processes in Lakes and Oceans*, ed. J. Imberger, Coastal and Estuarine Studies Volume 54, 19-36, American Geophysical Union.

Earle, M. D. (1996), Nondirectional and directional wave data analysis procedures, *NDBC Tech. Doc. 96-01* (http://www.ndbc.noaa.gov/wavemeas.pdf), 43 pp.

Ebuchi, N., and Kawamura, H. (1994), Validation of wind speeds and significant wave heights observed by the TOPEX altimeter around Japan, *J. Oceanography, 50*, 479--487.

Felizardo, F., and W. K. Melville (1995), Correlations between ambient noise and the ocean surface wave field, *J. Phys. Oceanogr., 25,* 513-532.

Freilich, M. H., and Challenor, P. G. (1994), A new approach for determining fully empirical altimeter wind speed model functions, *J. Geophys. Res*., *99*, 25051-25062.

Gemmrich, J. R., and D. M. Farmer (2004), Near-surface turbulence in the presence of breaking waves, *J. Phys. Oceanogr., 34,* 1067-1086.

Gemmrich, J. R., M. L. Banner, and C. Garrett (2008), Spectrally resolved energy dissipation rate and momentum flux of breaking waves, *J. Phys. Oceanogr., 38,* 1296-1312.

Gommenginger, C. P., M. A. Srokosz, P. G. Challenor and P. D. Cotton (2003), Measuring ocean wave period with satellite altimeters: A simple empirical model, *Geophys. Res. Lett., 30*, L222150, doi:10.1029/2003GL017743.

Gower, J. F. R. (1996), Intercomparison of wave and wind data from TOPEX/POSEIDON, *J. Geophys. Res*., *101*, 3817-3829.

Hanson, J. L., and O. M. Philips (1999), Wind sea growth and dissipation in the open ocean, *J. Phys. Oceanogr.*, *29*, 1633-1648.

Hasselmann, K., et al. (1973)**,** Measurements of wind-wave growth and swell decay during the Joint North Sea Wave Project (JONSWAP), *Deutsch. Hydrogra. Z., A8*, 95 pp.

Hwang, P. A. (2006), Duration- and fetch-limited growth functions of wind-generated waves parameterized with three different scaling wind velocities, *J. Geophys. Res*., *111*, C02005, doi:10.1029/2005JC003180.

Hwang, P. A., and M. A. Sletten (2008), Energy dissipation of wind-generated waves and whitecap coverage, *J. Geophys. Res*., *113*, C02012, doi:10.1029/2007JC004277.

Hwang, P. A., and D. W. Wang (2004), An empirical investigation of source term balance of small scale surface waves, *Geophys. Res. Lett*., *31*, L15301, doi:10.1029/2004GL020080.

Hwang, P. A., Teague, W. J., Jacobs, G. A., and Wang, D. W. (1998), A statistical comparison of wind speed, wave height and wave period derived from satellite altimeters and ocean buoys in the Gulf of Mexico Region, *J. Geophys. Res.*, *103*, 10451-10468.

Hwang, P. A., M. A. Sletten, and J. V. Toporkov (2008a), Analysis of radar sea return for breaking wave investigation, *J. Geophys. Res.*, *113*, C02003, doi:10.1029/2007JC004319.

Hwang, P. A., M. A. Sletten, and J. V. Toporkov (2008b), Breaking wave contribution to low grazing angle radar backscatter from the ocean surface, *J. Geophys. Res.*, *113*, C09017, doi:10.1029/2008JC004752.

IAHR (1989), List of sea state parameters, *J. Waterway Port Coast. Ocean. Eng.*, *115* , 793-808.

Phillips, O. M. (1977), *The dynamics of the upper ocean*, 2nd ed., Cambridge Univ. Press, Cambridge, UK, 336 pp.

Phillips, O. M. (1985), Spectral and statistical properties of the equilibrium range in wind-generated gravity waves, *J. Fluid Mech.*, *156*, 505-531.

Pierson, W. J., and L. Moskowitz (1964), A proposed spectral form for full, developed wind seas based on the similarity theory of S. A. Kitaigorodskii, *J. Geophys. Res.*, *69*, 5181-5190.

Terray, E. A., M. A. Donelan, Y. C. Agrawal, W. M. Drennan, K. K. Kahma, A. J. Williams, P. A. Hwang, and S. A. Kitaigorodskii (1996), Estimates of kinetic energy dissipation under breaking waves, *J. Phys. Oceanogr.*, *26*, 792-807.

Toba, Y. (1978), Stochastic form of the growth of wind waves in a single-parameter representation with physical interpretation, *J. Phys. Oceanogr.*, *8*, 494-507.

**Table 1. Buoy stations and satellite tracks in the four regions selected for this study.**

| NDBC Buoy ID | Buoy Location | TP Tracks | **Region** |
|---|---|---|---|
| **46001** | (56°17'44"N 148°10'19"W) | 11,27 | Gulf of Alaska |
| **46003** | (51°49'53"N 155°51'01"W) | 91, 100 | |
| **46035** | (56°54'38"N 177°48'38" W) | 28,101 | Bering Sea |
| **42002** | (25°53'30" N 93°34'03" W) | 26, 59 | Gulf of Mexico |
| **42003** | (25°56'10" N 85°54'51" W) | 46 | |
| **42020** | (27°00'44" N 96°30'20" W) | 115, 21 | |
| **42035** | (29°14'47" N 94°24'35" W) | 26 | |
| **42036** | (28°30'01" N 84°30'08" W) | 46 | |
| **51001** | (23°24'04"N 162°15'59" W) | 36,92 | Hawaiian Islands |
| **51002** | (17°10'12"N 157°48'24"W) | 3,23 | |
| **51003** | (19°10'17"N 160°43'47" W) | 92 | |
| **51004** | (17°26'12" N 152°31'10" W) | 79,99 | |
| **51028** | (00°00'03"N 153°51'28" W) | 92 | Equatorial Region |

**Table 2. Mean and standard deviation of wind and wave parameters in the four regions of this study.**

| Region | $U_{10}$ (m/s) | | $H_s$ (m) | | $T_p$ (s) | | $\omega_p U_{10}/g$ | | $T_a$ (s) | | $\omega_d U_{10}/g$ | |
|---|---|---|---|---|---|---|---|---|---|---|---|---|
| | Mean | Std. Dev. | Mean | Std. Dev. | Mean | Std. Dev. | Mean | Std. Dev. | Mean | Std. Dev. | Mean | Std. Dev. |
| **GOA** | 8.5 | 3.92 | 2.8 | 1.50 | 9.7 | 2.67 | 0.61 | 0.42 | 6.7 | 1.31 | 0.83 | 0.52 |
| **GOM** | 6.3 | 2.82 | 1.2 | 0.71 | 6.1 | 1.43 | 0.71 | 0.60 | 4.8 | 0.84 | 0.88 | 0.69 |
| **Hawaii** | 7.5 | 2.48 | 2.4 | 0.75 | 10.4 | 2.99 | 0.51 | 0.26 | 6.5 | 1.19 | 0.78 | 0.34 |
| **Equator** | 6.2 | 1.98 | 2.0 | 0.45 | 11.4 | 3.24 | 0.37 | 0.18 | 7.0 | 1.09 | 0.59 | 0.24 |

**Table 3. Statistics of comparison of the wave periods derived from altimeter and wave buoys (*B:* bias, *c:* orthogonal regression coefficient, *D:* rms difference, and *Q:* correlation coefficient).**

| Region | $B$ | $c$ | $D$ | $Q$ |
|---|---|---|---|---|
| **GOA** | -0.002 | 0.997 | 0.490 | 0.924 |
| **GOM** | 0.022 | 1.001 | 0.418 | 0.837 |
| **Hawaii** | 0.006 | 1.000 | 0.526 | 0.897 |
| **Equator** | 0.041 | 1.007 | 0.524 | 0.892 |

**Table 4. Comparison statistics of the average wave period derivation from the algorithms described in this paper and in *Gommenginger et al.* [2003] with the buoy measurement.**

| B | | c | | D | | Q | | $D_N$ | |
|---|---|---|---|---|---|---|---|---|---|
| H08 | G03 | H08 | G03 | H08 | G03 | H08 | G03 | H08 | G03 |
| -0.001 | 0.017 | 0.997 | 1.002 | 0.491 | 0.637 | 0.924 | 0.876 | 0.069 | 0.091 |
| 0.166 | -0.013 | 1.033 | 1.002 | 0.455 | 0.510 | 0.820 | 0.825 | 0.098 | 0.106 |
| 0.162 | 0.086 | 1.017 | 1.004 | 0.584 | 0.780 | 0.883 | 0.750 | 0.080 | 0.107 |
| -0.136 | -0.435 | 0.976 | 0.931 | 0.534 | 0.718 | 0.883 | 0.777 | 0.072 | 0.093 |

**List of figures**

Fig. 1. Similarity relation of wind speed, wave height and wave period expressed as (a) $\eta_*(\omega_{p*})$ and $\eta_*(\omega_{a*})$, (b) $U_{10}/gT_p$ and $U_{10}{}^2/gH_s$, and (c) $U_{10}/gT_a$ and $U_{10}{}^2/gH_s$.

Fig. 2. Map showing the buoy locations with black circles in the four geophysical regions discussed in this paper.

Fig. 3. The pdf of the wind and wave parameters in the four geophysical regions listed in Table 1: (a) $U_{10}$, (b) $H_s$, (c) $T_p$, (d) $\omega_{p*}$, (e) $T_a$, and (f) $\omega_{a*}$.

Fig. 4. The initial estimation of the wave period, $T_0$, derived with wind speed and wave height from altimeter using the algorithm of *Hwang et al.* [1998]. (a) Comparison with buoy $T_a$, illustrating the overestimation and wave age dependence. The numbers in the square brackets represent the range of wave age. (b) Comparison of the surrogate wave age from altimeter parameters and buoy wave age, and (c) the wave age trend to be removed to improve agreement between buoy and TP wave periods.

Fig. 5. The second and final estimations of the average wave period compared with buoy $T_a$: (a) $T_1$, and (b) $T_a$ from TP; (c) the much weaker wave age trend in the final result.

Fig. 6. Comparison of the average wave period derived with the algorithms described in this paper and in *Gommenginger et al.* [2003] for the four geophysical regions: (a, b) GOA, (c, d) GOM, (e, f) HAW, and (g, h) EQU. The top row is from the present algorithm, the bottom row is from the *Gommenginger et al.* algorithm.

Fig. 7. The horizontal energy flux coefficients: (a) $\alpha_h(\omega_{p*})$, and (b) $\alpha_h(\omega_{a*})$, calculated from the dimensionless wind and wave parameters: (c) $\eta_*(\omega_{p*})$, and (d) $\eta_*(\omega_{a*})$. Results from four geophysical regions are shown.

Fig. 8. (a) TP derived $\alpha_h(\omega_{aT*})$, (b) comparison of $\varepsilon_h(\omega_{a*})$ derived from TP and buoy; and the corresponding wind speed dependence: (c) $\alpha_h(U_{10})$ and (d) $\varepsilon_h(U_{10})$. Results from four geophysical regions are shown.

Fig. 9. Comparison of the vertical energy flux coefficient calculated with buoy and TP wind and wave parameters: (a) $\alpha_{vB}(\omega_{aB*})$, (b) $\alpha_{vT}(\omega_{aT*})$; and (c) wind speed dependence of the vertical energy flux. Results from four geophysical regions are shown.

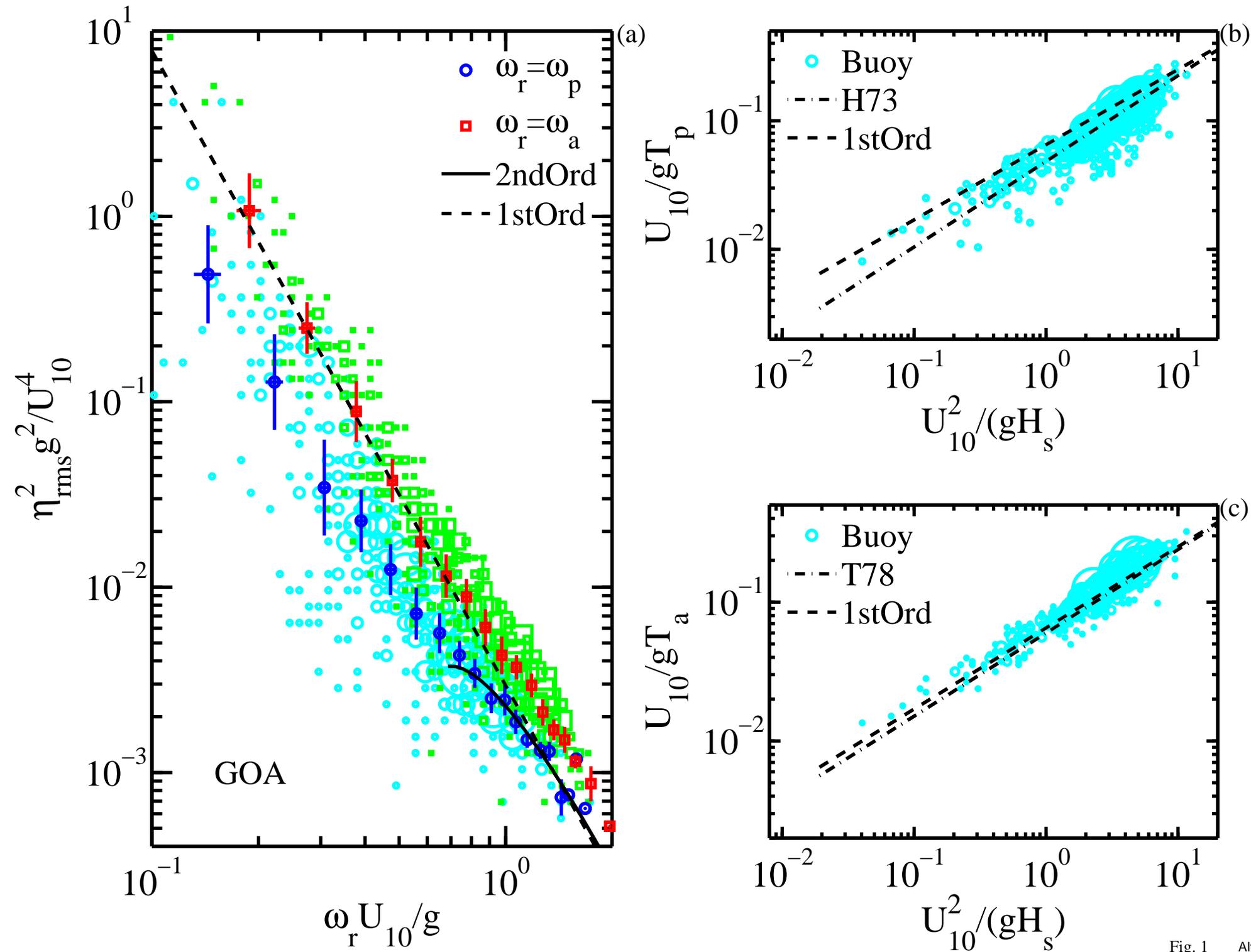

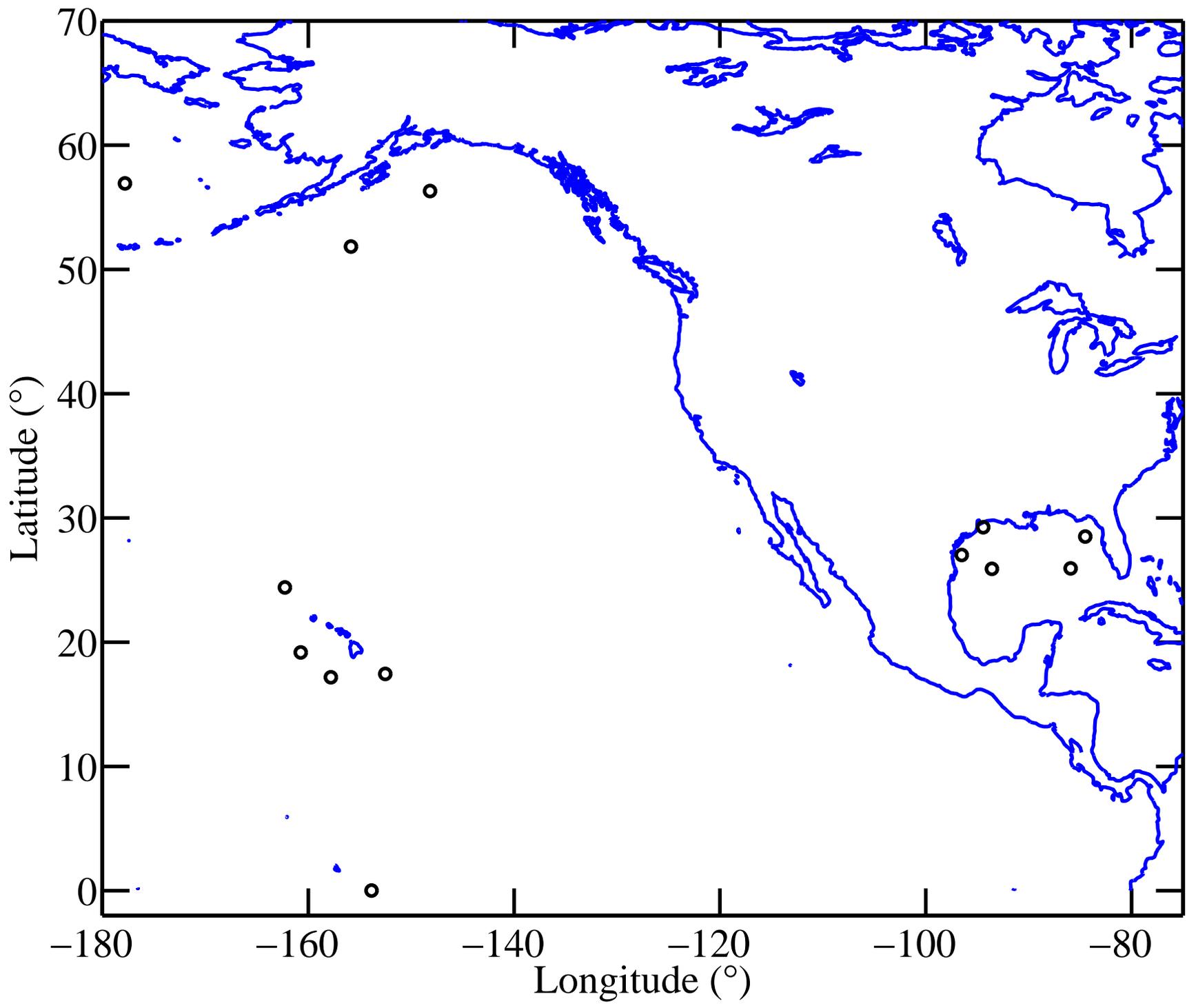

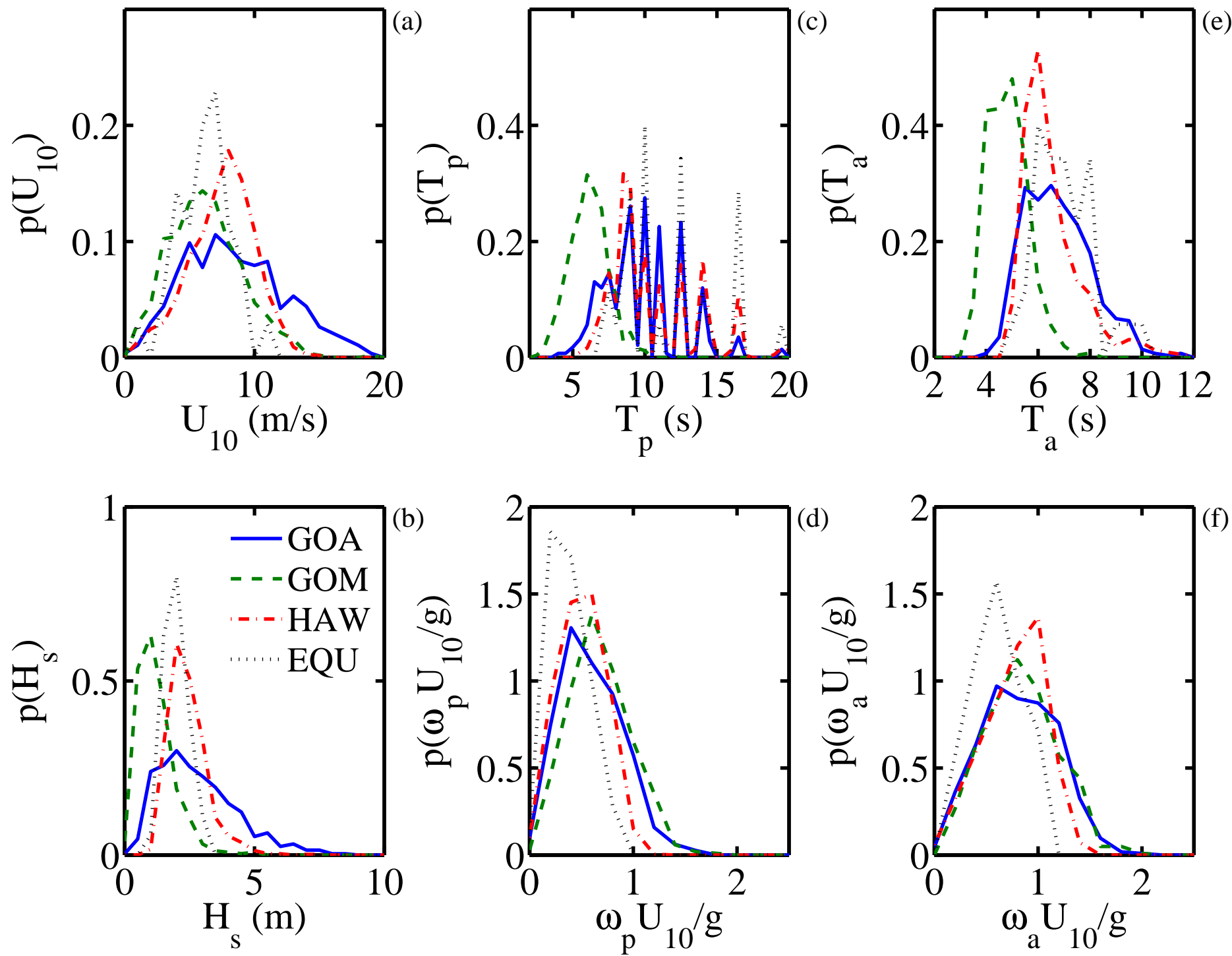

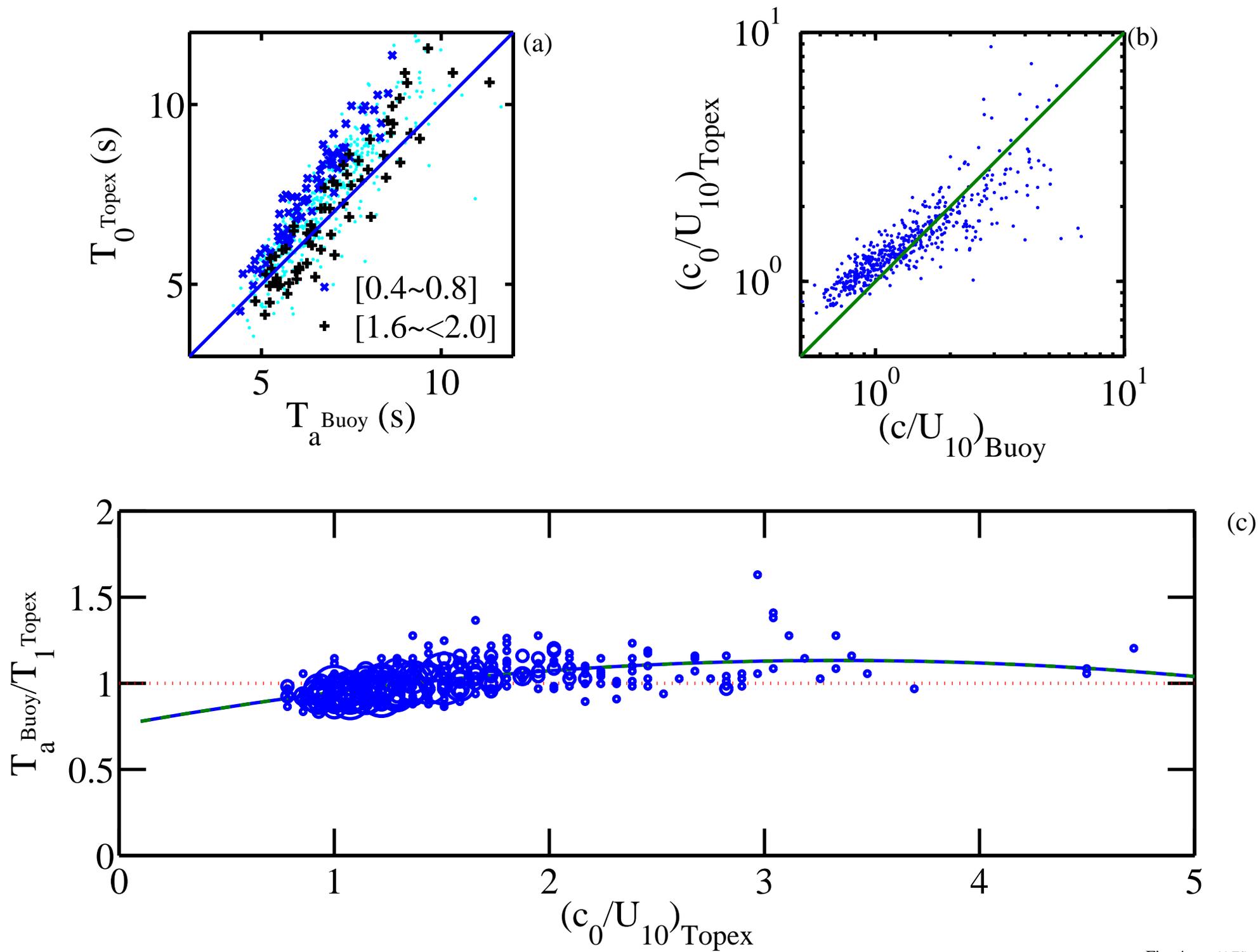

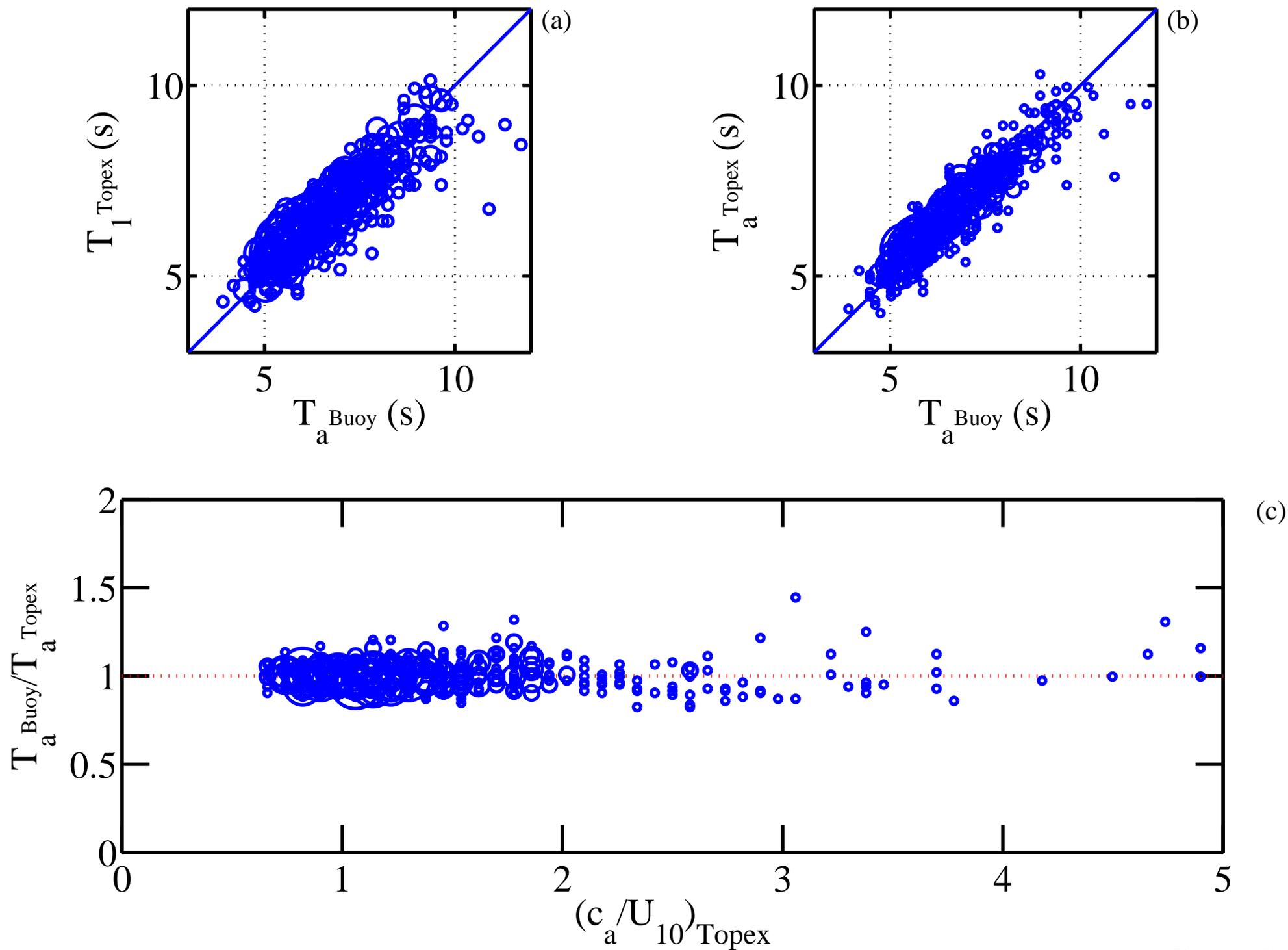

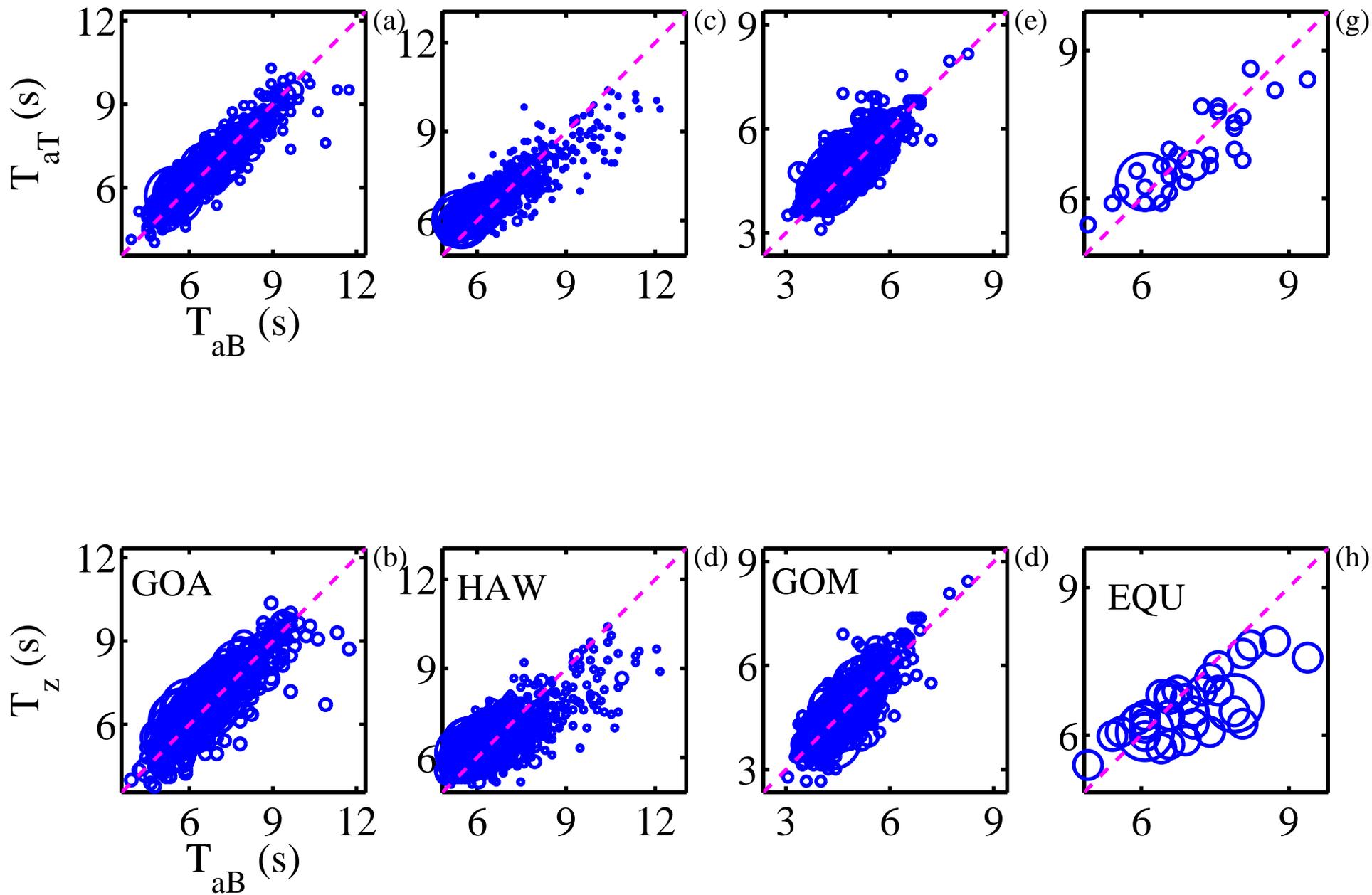

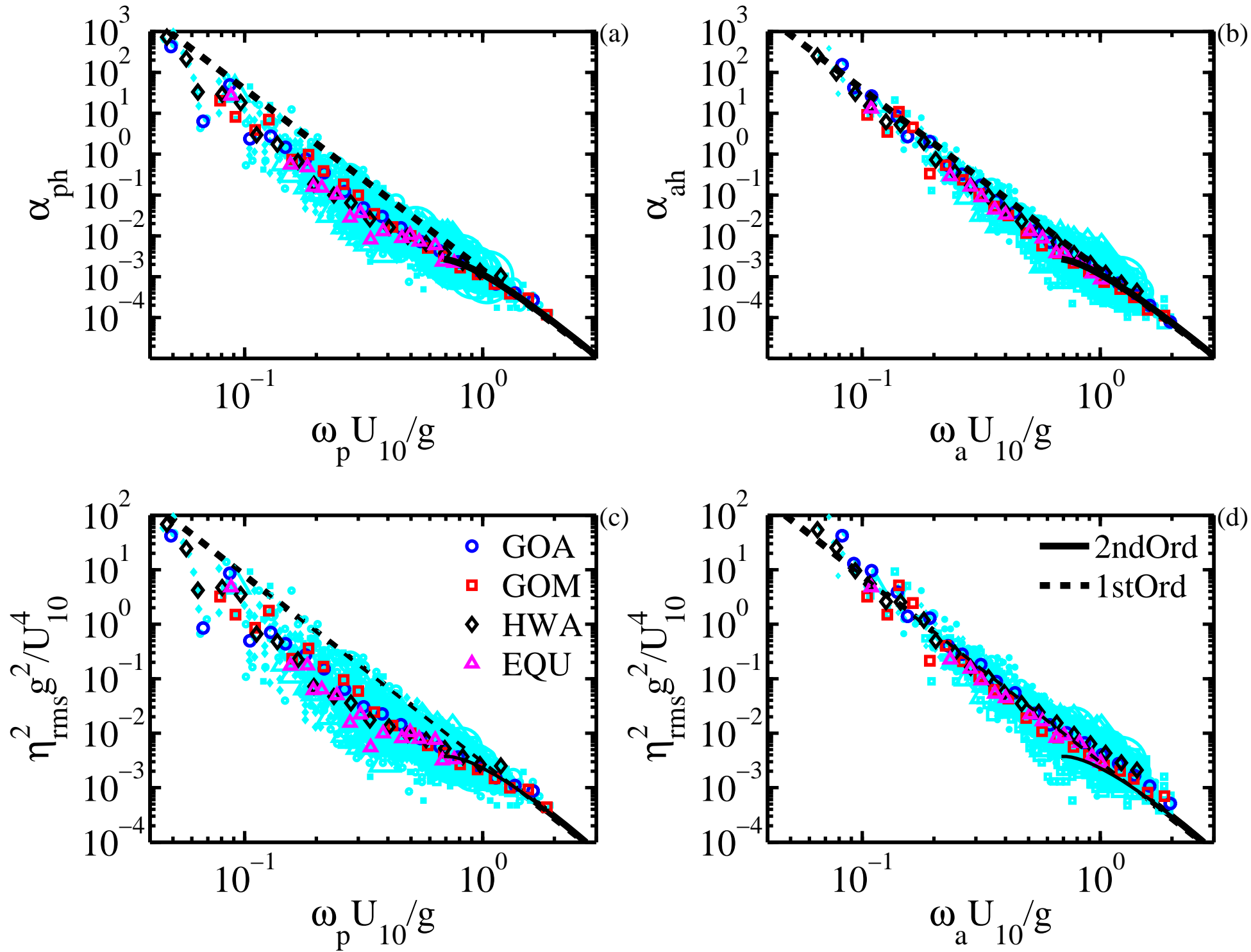

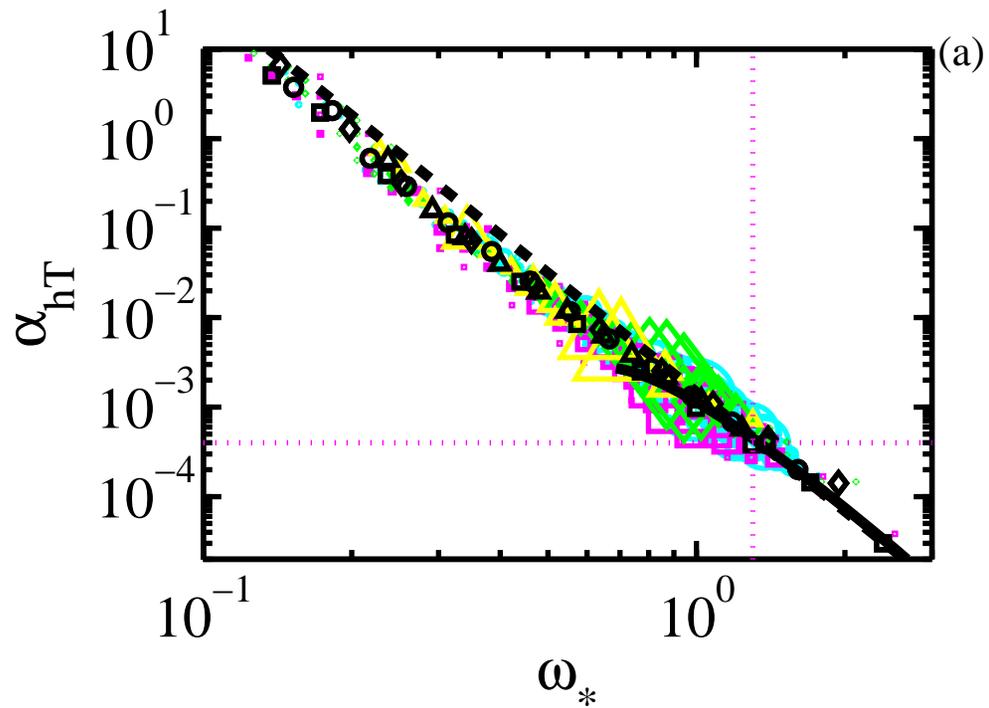

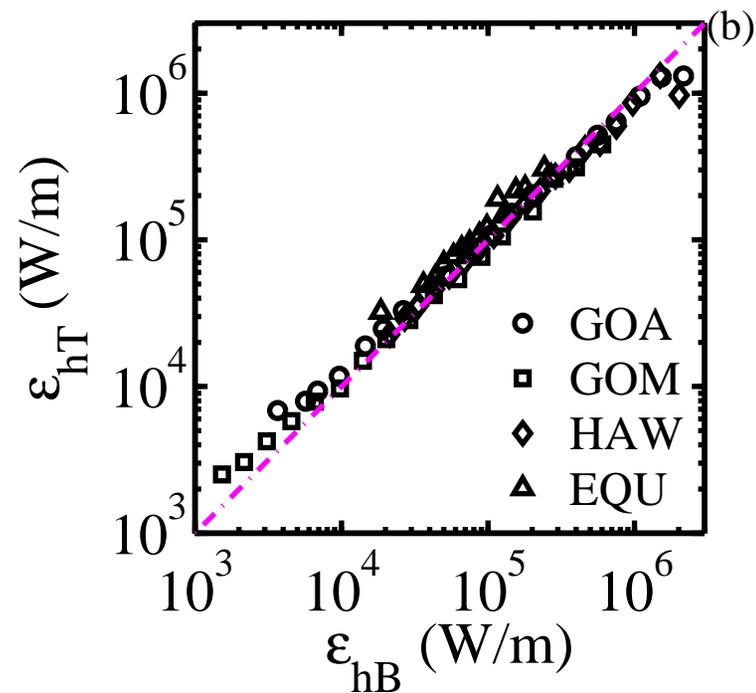

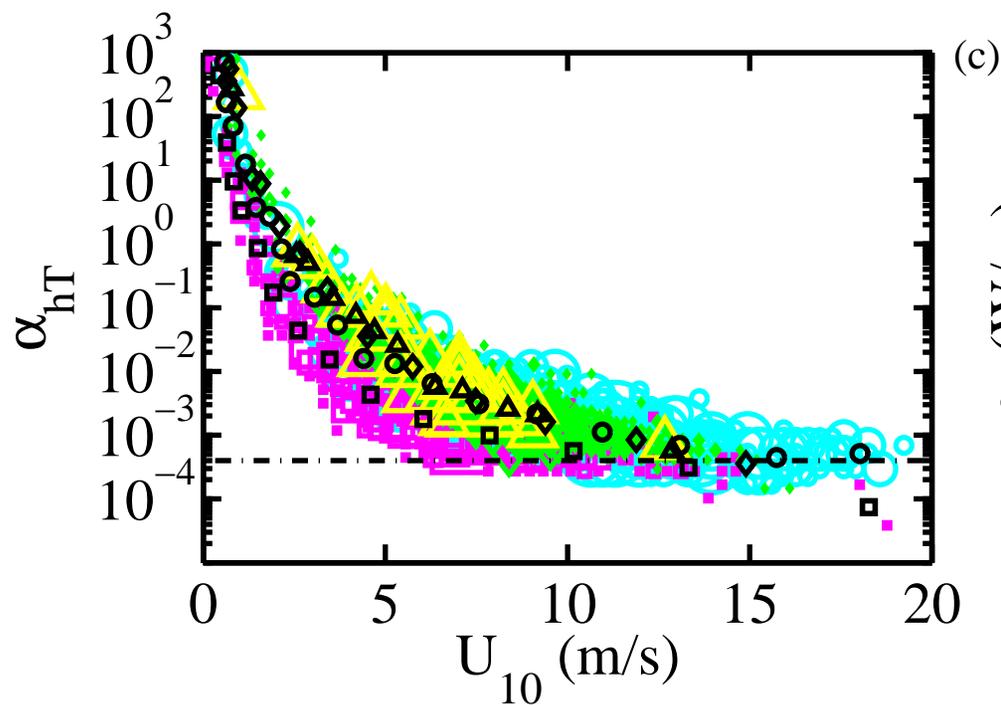

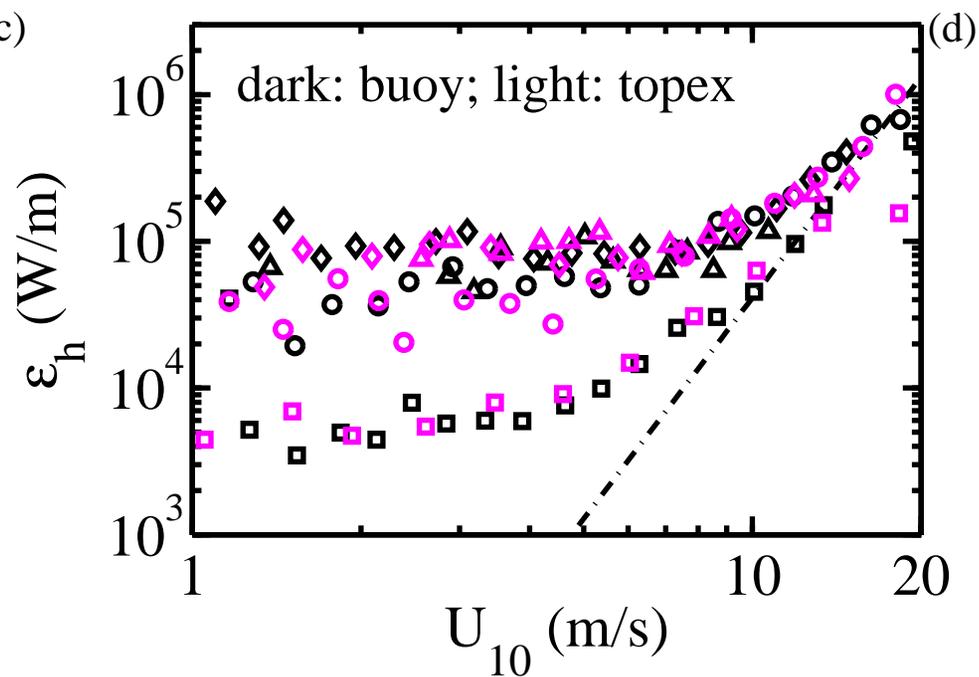

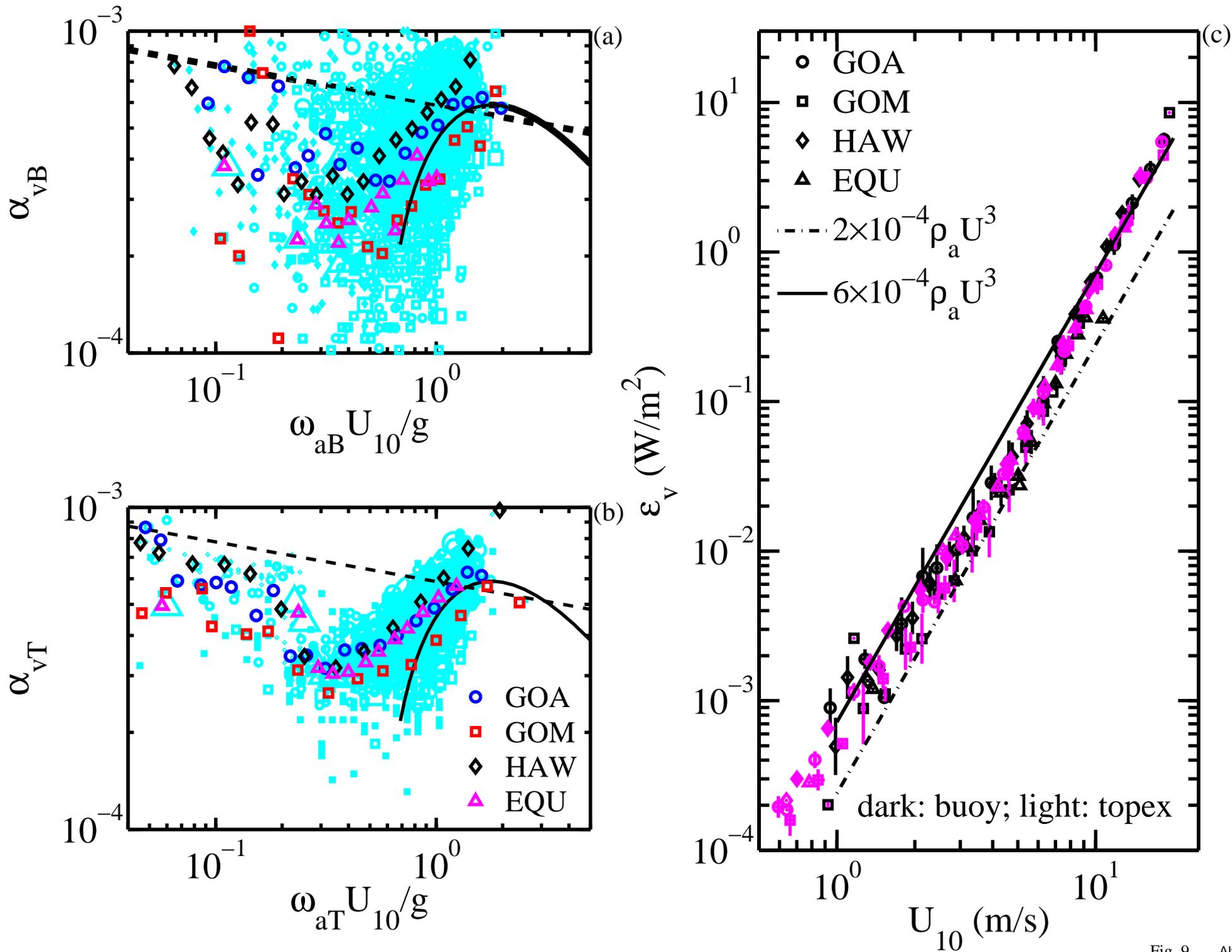

Fig. 9    AltEfluxFixR0